\documentclass[letterpaper, 10 pt, conference]{ieeeconf}  

\IEEEoverridecommandlockouts                              

\usepackage{graphics} 
\usepackage{epsfig} 
\usepackage{times} 
\usepackage{amsmath} 
\usepackage{amssymb}  
\usepackage{fancyhdr}
\usepackage{url}
\usepackage{hyperref}
\hypersetup{
	colorlinks=true,
	linkcolor=black,
	filecolor=magenta,      
	urlcolor=cyan,
}
\usepackage[dvipsnames]{xcolor}
\definecolor{antiquefuchsia}{rgb}{0.57, 0.36, 0.51}
\definecolor{DarkGreen}{rgb}{0.57, 0.8, 1}

\def \SO {{\mathbf{SO}}}
\def \so {{\mathfrak{so}}}

\newtheorem{thm}{Theorem}
\newtheorem{prop}{Proposition}

\newtheorem{lem}{Lemma}
\newtheorem{assumption}{Assumption}

\newtheorem{problem}{Problem}
\title{\LARGE \bf
	{Constructive barrier feedback for collision avoidance in leader-follower formation control
}}

\author{Zhiqi~Tang$^{1,2}$, Rita Cunha$^{1}$, Tarek Hamel$^{3}$, Carlos Silvestre$^{1,4}$
	\thanks{*This work was partially supported by the Project MYRG2015-00126-FST of the University
		of Macau; by the Macao Science and Technology, Development Fund under Grant FDCT/0031/2020/AFJ; by Funda\c{c}\~{a}o  para a Ci\^{e}ncia e a Tecnologia (FCT) through Project UIDP/50009/2020 and Project PTDC/EEI-AUT/1732/2020;
		and by the EQUIPEX project Robotex. }
	\thanks{$^1$Institute for Systems and Robotics, Instituto Superior T\'{e}cnico, Universidade de Lisboa, Portugal.}
	\thanks{$^2$Division of Decision and Control Systems, KTH Royal Institute of Technology, Sweden.}
	\thanks{$^3$I3S-CNRS, Universit\'{e} C\^{o}te d'Azur, and Institut Universitaire de France, France.}
	\thanks{$^4$Department of Electrical and Computer Engineering of the Faculty of Science and Technology of the University of Macau, Macao, China.}%
	\thanks{{\tt\small Email: zhiqitang@tecnico.ulisboa.pt, rita@isr.tecnico.ulisboa.pt, thamel@i3s.unice.fr, csilvestre@umac.mo}}
}

\begin{document}
	
	\maketitle
	\cfoot{\thepage}
	\begin{abstract}
This paper proposes a novel constructive barrier feedback for reactive collision avoidance between two agents. It incorporates this feature in a formation tracking control strategy for a group of 2nd-order dynamic robots defined in three-dimensional space. Using only relative measurements between neighboring agents, we propose an elegant decentralized controller as the sum of a nominal tracking controller and the constructive barrier feedback for leader-follower formations under a directed single-spanning tree graph topology. The key ingredient is the use of divergent flow as a dissipative term, which slows down the relative velocity in the direction of the neighboring robots without compromising the nominal controller’s performance. Compared to traditional barrier function-based optimization controllers, the proposed constructive barrier feedback avoids feasibility issues and results in more computationally efficient control algorithms with systematic equilibrium analysis.
Simulations are provided to validate the effectiveness of the proposed control method. 
 

	\end{abstract}

	\section{INTRODUCTION}
	The topic of multi-robot coordination has garnered increasing interest in both robotics and control communities owing to its vast potential in various applications, such as infrastructure inspection, intelligent transportation, and exploration of unknown environments. A multi-robot system's primary goal in a mission is typically to solve a task collaboratively  while maintaining a certain relative position between neighboring agents. 
	Collision avoidance, typically considered a secondary objective, is critical for safety. 
 
 The field of multi-robot formation control with collision avoidance has seen significant advancements over the years. Early works mainly focused on designing gradient descent control laws based on constructing the potential function using geometric information on the considered topology \cite{rimon1990exact, kan2011network}. Although convergence to the desired and undesired equilibrium points is explicitly analyzed, these solutions were limited to multi-robot systems under single-integrator systems operating in two-dimensional space.
	More recently, optimization-based controllers that employ control barrier functions have emerged as promising alternatives (e.g., \cite{wang2017safety, reis2020control}) to guarantee collision-free behavior between robots.
	In \cite{wang2017safety}, a decentralized safety controller is proposed for multi-robot systems modeled as second-order systems by solving a quadratic programming problem. In contrast, the work  in \cite{wang2017safety} focuses on analyzing the conditions that guarantee a safe distance between robots without explicitly analyzing the system's equilibrium. It is worth noting that the use of barrier function-based optimization controllers poses challenges in explicitly analyzing the equilibrium and convergence of the multi-robot system, in addition to potential computational complexity and feasibility issues. As pointed out in \cite{reis2020control}, even in a simple scenario where only one single-integrator robot with one static obstacle is considered, undesirable asymptotically stable equilibria can exist on the boundary of the safety set. 
	
This paper introduces a novel control method inspired by natural systems like insects and birds, which exploits \textit{divergent flow} \cite{bhagavatula2011optic} to prevent collisions while effectively achieving the primary control objective. Previous research has demonstrated divergent flow's capacity for ensuring safe quadrotor landings in cluttered environments \cite{rosa2014optical} and on moving platforms \cite{herisse12}, as well as its application in avoiding collisions in flocking models \cite{Herisse23}. While promising, the full potential of divergent flow for collision avoidance remains unexplored.

	This paper exploits the divergent flow to achieve safe collision avoidance without compromising the primary control objective, providing a robust and intuitive solution for multi-agent systems in cluttered environments.
In particular, we address collision-free formation control design for second-order dynamic robots in three-dimensional space. The proposed method adds the divergent flow directly to a nominal leader-follower formation tracking controller that stabilizes the formation tracking error. The added divergent flow acts as a dissipative function, slowing down the relative velocity in the direction of neighboring robots without affecting the nominal controller's performance. By assuming the formation has a directed single-spanning tree graph topology with a leader controlled independently, we guarantee that 1) there is no collision between neighboring agents, and 2) the configuration of the formation either converges to the asymptotically to the desired configuration or a finite number of unstable set points. To the best of the authors' knowledge, this is the first work that exploits divergent flow in formation tracking control for collision avoidance with systematic equilibria analyses.

	The remainder of this paper is organized into four sections. Section II provides notation and some mathematical preliminaries on graph theory. Section III presents the proposed control method with stability analysis. Simulation results are presented in Section IV, and some final comments are discussed in Section V.

	\section{Preliminaries on graph theory}
	Let $\mathbb{S}^2:=\{y\in\mathbb{R}^3:\|y\|=1\}$ denote the 2-Sphere and $\|.\|$ the Euclidean norm. For any $y\in \mathbb{S}^2$, we define the projection operator $\pi_y$
	\begin{align*}
		\pi_y := I - y y^{\top} \geq 0, 
	\end{align*}
	such that, for any vector $x\in\mathbb{R}^3$, $\pi_y x$ provides the projection of $x$ on the plane orthogonal to $y$. For any $z\in\mathbb{R}^3$, $z_\times$ represents the skew-symmetric matrix associated with the vector $z$. Let $\SO(3):= \{
	R \in \mathbb{R}^{3 \times 3} \mid R^\top R = I_3, \; \det(R) = 1
	\}$ be the special orthogonal group and the Lie algebra $\so(3):=\{
	\Omega \in \mathbb{R}^{3 \times 3} \mid \Omega ^\top=-\Omega\}$ be the set of skew-symmetric
	$3 \times 3$ matrices, such that $\dot{R}=R\Omega$.

	Consider a system of $n\ (n\ge 2)$ connected agents. The underlying interaction topology can be modeled as a digraph (directed graph) $\mathcal{G}:= (\mathcal{V}, \mathcal{E})$, where $\mathcal{V}=\{1,2,\ldots,n\}$ is the set of vertices and $\mathcal{E} \subseteq \mathcal{V} \times \mathcal{V}$ is the set of directed edges.  The set of neighbors of agent $i$ is denoted by $\mathcal{N}_i:=\{j\in\mathcal{V}|(i,j)\in\mathcal{E}\}$. 
	To provide clarity on the graph topology used in this work, we make the following assumption:
	\begin{assumption}\label{ass:topology}
		The topology $\mathcal G$ is fixed and described by an acyclic digraph with a single directed spanning tree.
		Without loss of generality, agents are numbered (or can be renumbered) such that agent $1$ is the leader, i.e.  $\mathcal{N}_1= \varnothing$,  all other agents $i, \ i\ge 2$ are followers whose neighboring set is $\mathcal{N}_i = \{i-1\}$. 
	\end{assumption}

	\section{Reactive collision avoidance during leader-follower formation tracking control}
	
	In this section, we consider a constructive control design for the leader-follower formation tracking control problem with reactive inter-agent collision avoidance. Given a directed graph $\mathcal{G}$, we denote the position of each agent $i \in \mathcal{V}$ expressed in a common inertial frame as $p_i \in \mathbb{R}^3$. The stacked vector $\boldsymbol{p}=[p_1^\top,...,p_n^\top]^\top\in \mathbb{R}^{3n}$ is defined as the configuration of $\mathcal{G}$, and the formation $\mathcal{G}(\boldsymbol p)$ in 3-dimensional space is determined by $\mathcal{G}$ and $\boldsymbol{p}$.
	
	We consider the case in which a double integrator describes the individual robot dynamics. That is:
	\begin{equation}\label{eq:double integrator}
		\left\{
		\begin{aligned}
			\dot{p}_i&=v_i\\
			\dot{v}_i&=u_i
		\end{aligned}
		\right.
	\end{equation}
	where $v_i\in\mathbb{R}^3$ is velocity of each agent $i$ and and $u_i\in\mathbb{R}^3$ its acceleration used here as the control input. 
	\begin{figure}[!htb]
		\centering
		\includegraphics[scale = 0.6]{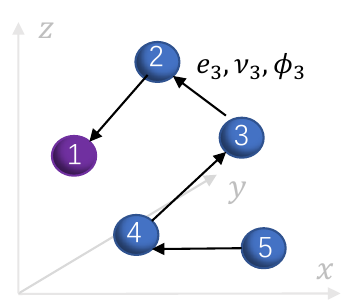}	
		\caption{Interaction of a 5-agent leader-follower formation $\mathcal G(\boldsymbol p)$ in three-dimensional space. The arrows indicate each agent $i, i\ge2 $ can measure the relative states to its neighboring agent $i-1$: $e_i$, $\nu_i$, and $\phi_i$. }
		\label{fig:graph}
	\end{figure}
	
	In the considered scenario, the topology of the formation $\mathcal G (\boldsymbol p)$ satisfies Assumption \ref{ass:topology}
	such that each agent $i$ ($i = 2, \ldots, n$) only has access to information relative to its topological neighbor $i-1$ as illustrated in Fig. \ref{fig:graph}. 
	To simplify the notation, we define the relative position vectors between two neighboring agents $i$ and $i-1$ as:
	\begin{equation}
		\label{eq:eij}
		e_{i}:=p_{i}-p_{i-1}, \ i\ge 2.
	\end{equation}
	Similarly, $\nu_{i}:=\dot e_{i}=v_i-v_{i-1}$ denotes the relative velocity between agent $i$ and $i-1$. As long as $\|e_{i}\|\ne 0$, one can define  direction vector from $i$ to $i-1$ as:
	$$g_{i}=\frac{e_{i}}{\|e_{i}\|}.$$
	
	Let $r$ be a positive constant that we term the safety distance and define $d_{i}:=\|e_{i}\|-r=\|p_i-p_{i-1}\|-r$. A straightforward computation shows that $\dot d_{i}=g_{i}^\top \nu_{i}$.
	
	The desired configuration $\boldsymbol p^*$ of $\mathcal{G}(\boldsymbol p^*)$ is assumed to be smooth in 3-dimensional space such that for some constant $D>r$,  $r<|e^*_{i}|<D$.
	The control objective considered is formally described in the following.
	
	\begin{problem}\label{problem}
		Find individual feedback control actions $u_i$ depending only on the relative measured states and the first and second derivatives of the smooth trajectory $p^*_i$ such that the tracking error $(\boldsymbol p-\boldsymbol p^*)$
		converges towards zero for all bounded initial conditions such that $\|e_{i}(0)\|>r$ while guaranteeing that $\|e_{i}\|>r$ (resp. $d_{i}>0$) for all time.
	\end{problem}
	
	Without a precise description of the measures to be used as feedback information and making additional assumptions on the desired trajectory, the difficulty level of solving the problem as described is high.
	The relative states $e_{i}$ and $\nu_{i}$ are the natural information to be used in a nominal control design to ensure asymptotic (or exponential) stability of the equilibrium $(e_i-e_i^*,\nu_i-\nu_i^*)=(0,0)$. To prevent collisions between neighboring agents, the key principle is controlling the relative velocity along the direction $g_{i}$, i.e., $\dot d_{i}=g_{i}^\top \nu_{i}$. To get an effective reactive collision avoidance without affecting the nominal controller stability property, we design controller $u_i$ as:
	\begin{equation}\label{eq:ui}
		u_i=u_i^n-k_{o}g_{i}f^B(\dot d_{i}, d_{i}), \ i\in \mathcal V/\{1\},
	\end{equation}
	with $k_{o}$ a positive constant gain and $u_i^n=u_i^n(\tilde{e}_{i},\tilde{\nu}_{i})$ the nominal control input ensuring the asymptotic (or the exponential) stability of the equilibrium $(e_i-e_i^*,\nu_i-\nu_i^*)=(0,0)$  and $f^B$ a dissipative control barrier feedback slowing down the relative velocity in the direction of the neighbor without compromising the stability nature of the nominal control action. 
	
	To the best of our knowledge, in most applications, the divergent flow is the simplest and most elegant choice for $f^B$. That is: 
	\begin{equation}\label{f}
		\phi_i:=f^B(\dot d_{i}, d_{i})=\frac{\dot d_{i}}{d_{i}}=\frac{g_{i}^\top \nu_{i}}{d_{i}}.
	\end{equation}
	It can be obtained directly from the optical flow using visual information \cite{rosa2014optical}, built or estimated from the measure of $d_{i}$ \cite{hua2010obstacle}.

	To illustrate the obstacle avoidance principle employed in this context, let's consider a 2-agent system with agent $i-1$ as the leader, agent $i$ as the follower.   Using the above definitions of $d=d_{i}$ and $\dot{d}=\dot d_{i}$, it is straightforward to verify that:
	\begin{equation} \label{ddot_d}
		\ddot d=-k_o\frac{\dot d }{d} -\alpha(t) 
	\end{equation}
	with $\alpha(t)=-\frac{\|\pi_{g_{i}}\nu_{i}\|^2}{d+r} +g_{i} ^\top({u}^n_{i}-u_{i-1})$.
	The barrier effect of the $f^B$, is announced  in the following technical Lemma:
	\begin{lem}\label{lem:boundness of OF}
			Given the dynamics \eqref{ddot_d}
			with $k_o$ a positive gain and $\alpha(t)$ a continuous and bounded function. Then for any initial condition satisfying $d(0)>0$ and $\phi(0)=\frac{\dot d(0)} {d(0)}$ bounded, the following assertions hold:
			\begin{enumerate}
				\item $d$ remains positive, $\forall t\ge 0$.
				\item $d$ converges to zero as $t\to \infty$ if and only if (iff) $\lim_{t \to \infty} \int^t_0 \alpha(\tau) d\tau \to +\infty$. 
				\item If $d$ converges to zero, then $\dot{d}$ is bounded and converges to zero, and $\phi(t)$ remains bounded, $\forall t\ge 0$. Furthermore, if $\alpha(t)$ converges to a positive constant $\alpha^0>\epsilon >0$, then $\frac{\dot d}{d}\to -\frac{\alpha^0}{k_o}$ and hence $\ddot d$ converges to zero.
			\end{enumerate}
	\end{lem}

	Proof of the Lemma is given in Appendix \ref{app:L1}. This Lemma shows that $d=d_i$ will never cross zero for all the time as long as the baseline controller ${u}^n_i$, the leader input $u_{i-1}$, and the relative velocity $\nu_i$ are  continuous and bounded. 
	
	The control architecture adopted here differs from classical barrier function approaches by incorporating explicit control barrier feedback instead of relying on quadratic programming barrier constraints. Unlike traditional potential field methods, the proposed approach utilizes control barrier feedback for damping rather than repulsive effects, eliminating local minimal issues.

	Note, however, that by involving a simple additive control barrier feedback in the control design to avoid collisions between agents according to the graph topology while ensuring asymptotic (or exponential) stability of the
equilibrium, the following assumption on the desired formation is required, which involves the desired relative positions $e_{i}^*=p_i^*-p_{i-1}^*,\ i\ge 2$.
	\begin{assumption}\label{ass:desired} 
		The desired trajectories $\boldsymbol p^*$ of the formation are smooth 
		such that: the edges satisfy $e^*_i = c_i g_i^*, i\ge 2$, where $\|g^*_i\|=1$ and $c_i$ is a constant larger than $r$
	\end{assumption}	
	This assumption implied that $e_i^*$ is subjected to a distance-preserving action and $\dot{e}_{i}^*(t)=\Omega_{i}^{*} e_{i}^*$ with $\Omega_{i}^* \in \so(3)$ any smooth and bounded signal.  When $\Omega_{i}^*=\Omega^*, \; \forall i\in\mathcal V$, one verifies that the desired formation is subject to a rigid motion transformation for which $\Omega^*=\dot{R}^*R^{*\top}$, with $R^*\in SO(3)$ the rotation of the desired formation with respect to a fixed reference frame. When $\Omega_{i}^*=\Omega^*=0$, one has a rigid desired formation with only translational motion. From there, one defines the desired relative velocity and acceleration as follows:
	\begin{align*}
		\nu_{i}^*(e_{i}^*) &:= v_i^*-v_{i-1}^* = \Omega_{i}^{*}  e_{i}^*\\
		u_{e_i}^*(e_{i}^*) &:= u_i^*-u_{i-1}^* =\left( \dot\Omega_{i}^*+\Omega_{i}^{*^2}\right)e_{i}^*
	\end{align*}
	\begin{assumption}\label{assumption-leader}
		The leader is independently controlled from the formation such that one assumes without loss of generality that $p_1=p_1^*$, $v_1=v_1^*$ and $u_1=u_1^*$.
	\end{assumption}
	From Assumptions \ref{ass:desired} and \ref{assumption-leader}, one verifies that the  desired absolute position for each agent $i\ge 2$ can be explicitly expressed as:
	\begin{equation}\label{ref:pd}
		p_i^* = p_1^* + \sum_{j=2}^{i} c_j g_j^*.
	\end{equation}

	\subsection{Formation tracking control design and stability analysis}
	Define $\tilde{e}_{i}:=e_{i}-e_{i}^*$ and taking the time derivative of $\tilde{e}_{i}$, one gets:
	\begin{equation*}
		\dot{\tilde{e}}_{i}=\nu_{i}-\nu_{i}^*.
	\end{equation*}
	Using the fact that $\nu_{i}^*=\Omega_{i}^* (e_{i}-\tilde{e}_{i})$, 
one verifies that:
\begin{equation*}
	\dot{\tilde{e}}_{i}=\Omega_{i}^* \tilde{e}_{i} +\tilde{\nu}_{i}
\end{equation*}
where $\tilde{\nu}_{i}:=\nu_{i}- \nu_{i}^*(e_i)$ and $\nu_{i}^*(e_i):=\Omega_{i}^* e_{i}$ which is different from $\nu_{i}^*(e_i^*)$. With this choice, one verifies that $\dot d_{i}=g_{i}^\top \nu_{i}=g_{i}^\top \tilde{\nu}_{i}$ which in turn implies that the divergent flow is independent of the desired trajectory.

Taking now the time derivative of $\tilde{\nu}_{i}$ yields:
\begin{align*}
	\dot{\tilde{\nu}}_{i}&=-\Omega_{i}^* \tilde{\nu}_{i}+u_{e_i}-u_{e_i}^*(e_{i})\\
	&= -\Omega_{i}^* \tilde{\nu}_{i}+u_{i}-u_{i-1}-u_{e_i}^*(e_{i})
\end{align*}
with $u_{e_i}^*(e_{i}):=\left( \dot\Omega_{i}^*+\Omega_{i}^{*^2}\right)e_{i}$.
To achieve asymptotic tracking of the desired formation by each follower $i\ge 2$ under Problem \ref{problem} and Assumption \ref{ass:desired}, we suggest using a PD-like controller for the nominal control input:
\begin{equation}\label{eq:u_n}\scalebox{1}{$
		u_i^n=-k_{p}h_{p_i}(.)\tilde e_{i}-k_vh_{v_i}(.)\tilde \nu_{i}+u_{e_i}^*(e_{i})+u_{i-1},
		$}
\end{equation}
with $k_{p}$, $k_v$ two positive scalar gains.
The scalar functions $h_{p_i}(.)$ and $h_{v_i}(.)$ are smooth and bounded strictly positive functions defined on $[0,\infty)$ such that for some positive constant $\epsilon, \ \eta$ and $\beta$
\begin{align}\label{eq:hs}
	\forall s\in \mathbb R, \ \epsilon <h(s)< \eta, \ \ 0<|\frac{\partial h(s)}{\partial s}|< \beta.
\end{align}
An example of such a function is $h:s\mapsto h(s)=\eta/\sqrt{1+s}$ and a possible choice of $h_{p_i}(.)$ and $h_{v_i}(.)$ is  $h_{p_i}(.)=h(\|\tilde e_{i}\|^2)$ and $h_{v_i}(.)=h(\|\tilde \nu_{i}\|)$.

Recalling \eqref{eq:double integrator}, \eqref{eq:ui} and \eqref{eq:u_n}, one verifies that the closed-loop dynamics of the error variable $(\tilde e_{i}, \tilde \nu_{i}),i\in \mathcal V / \{1\}$ can be rewritten as
\begin{equation}\scalebox{0.9}{$ \label{eq:states_cl}
		\left\{
		\begin{aligned}
			\dot{\tilde{e}}_{i}=&\Omega_{{i}}^*\tilde e_{i}+\tilde {\nu}_{i}\\
			\dot{\tilde {\nu}}_{i}=&-\Omega_{{i}}^*\tilde \nu_{i}-k_{p}h_{p_i}(.)\tilde e_{i}-k_vh_{v_i}(.)\tilde \nu_{i}-k_{o}g_{i}\phi_i
		\end{aligned}
		\right.$}
\end{equation}
The following lemma provides the stability analysis of the first follower under the proposed controller.
\begin{lem} \label{lem:2agent}
	Consider a $2$-agent system with the dynamics \eqref{eq:double integrator} and let the input $u_2$ be given by \eqref{eq:ui} along with \eqref{f} and \eqref{eq:u_n}. 
	If Assumptions \ref{ass:topology}-\ref{assumption-leader} are satisfied, then for any initial conditions $(\tilde e_2(0),\tilde \nu_2(0))$ such that $d_2(0)>0$ and $\phi_2(0)$ is bounded, the following assertions hold
	\begin{enumerate}
		\item $d_2(t)$ remains positive and $\phi_2(t)$ and the control input $u_2$ are bounded $\forall t\ge 0$;
		\item $(\tilde{e}_{2},\tilde{\nu}_{2})$ converges either to the asymptotically  stable (AS) equilibrium point $(0,0)$, or to the unstable set point $(-(r+c_2) g_2^*, 0)$.
	\end{enumerate}

\end{lem}
	The proof of this Lemma is given in Appendix \ref{app:L2}.
	
By assuming that any follower $i, i\in\mathcal V/\{1\}$ in the $\mathcal G (\boldsymbol p)$ formation has access to $u_{i-1}$ in addition to the relative measures $e_i$, $\nu_i$, and $\phi_i$, the stability analysis presented in Lemma \ref{lem:2agent} can be extended to cover the trajectory tracking of any formation including more than two agents (i.e., $n>2$). Let $\tilde p_i$ and $\tilde v_i$ be the absolute errors: $\tilde p_i:=p_i-p_i^*$ and $\tilde v_i:=v_i-v_i^*, \ \forall i\in \mathcal V/ \{1\}$.  We will show the collision avoidance ability and analyze the convergence of the formation of $n\ge 2$ agents in the following Theorem.
\begin{thm}\label{thm:control_c}
	Consider an $n$-agent ($n\ge 2$) system with the dynamics \eqref{eq:double integrator} along with the feedback control law \eqref{eq:ui} and \eqref{eq:u_n}. 
	If Assumptions \ref{ass:topology}-\ref{assumption-leader} are satisfied, then for any bounded initial conditions $(\tilde e_i(0),\tilde \nu_i(0))$ such that $d_{i}(0)>0$ and $\phi_i(0)$ is bounded, then $\forall i\in \mathcal V/\{1\}$, the following assertions hold:
	\begin{enumerate}
		\item $d_i(t)$ remains positive and $\phi_i(t)$ and the control input $u_i$ are bounded $\forall t\ge 0$;
		\item $(\tilde{e}_{i},\tilde{\nu}_{i})$ converges either to the asymptotically  stable (AS) equilibrium point $(0,0)$ or to the unstable set point $(-(r+c_i) g_i^*, 0)$.
		\item The absolute state error $(\tilde p_i,\tilde v_i)$ converges to one of $2^{i-1}$ set points $(\bar {p}_i^m-p_i^*,0), m\in \{1,\ldots, 2^{i-1}\}$, where $\bar {p}_i^m = p_1^* + \sum_{\jmath=2}^{i} \mathring{c}_{j} g_{j}^*$ with $\mathring c_{j}\in\{c_{j},-r\}$  and $\bar{p}_i^1=p_{i}^*$, given by \eqref{ref:pd}. The solution $m=1$ leads to the unique asymptotically stable point $(\tilde p_i,\tilde v_i)=(0,0)$ and the remaining $2^{i-1}-1$ ($m\in \{2,\ldots, 2^{i-1}\}$) solutions are unstable. 
	\end{enumerate}
\end{thm}
The proof of this Theorem is provided in Appendix \ref{app:T1}.

		\subsection{Fully distributed formation tracking control design and stability analysis}
		Recall that the nominal controller \eqref{eq:u_n} involved in Theorem \ref{thm:control_c}, requires that each follower $i, i\ge 2$ have access to the actual input of its neighboring agent $u_{i-1}$ through communication. This assumption is very limiting in practical scenarios. To define a fully distributed controller in which each agent has access only to relative error to its neighbor, we redesign the nominal control law as follows $\forall i\ge 2$: 
		\begin{equation}\label{eq:u_nd}\scalebox{1}{$
				u_i^n=-k_{p}h_{p_i}(.)\tilde e_i-k_vh_{v_i}(.)\tilde \nu_i+u_{e_i}^*(e_i)+u_{i-1}^*.
				$}
		\end{equation}
		$u_{i-1}^*$ is iteratively obtained as follows: 
		\[u_{i-1}^*=u_{i-2}^* + u_{e_{i-1}}^*(e_{i-1}^*), i\ge 3,   \quad u_1^*=\ddot{p}_1^*. \]
		Note that when $\Omega_i^*=\Omega^*=0, \forall i\ge 2$, one verifies that $u_{e_i}^*(e_i)=u_{e_i}^*(e_i^*)=0$ and $u_i^*=u_1^*$, which is the desired linear acceleration of each agent. Compared to the previous nominal controller \eqref{eq:u_n}, the modified one \eqref{eq:u_nd} involves the desired acceleration $u_{i-1}^*$ of the neighboring agent instead of its actual acceleration $u_{i-1}$ introducing a cascaded structure into the system. The stability analysis of the system under the new nominal controller is provided in the following Proposition.
		\begin{prop}\label{thm:control}
			Consider the $n$-agent system with conditions described in Theorem \ref{thm:control_c}. Use the new nominal controller \eqref{eq:u_nd} instead of \eqref{eq:u_n} in the applied control \eqref{eq:ui}.  
			Then for any bounded initial conditions $(\tilde e_i(0),\tilde \nu_i(0))$ satisfying $d_{i}(0)>0$ and $\phi_i(0)$ bounded, the following assertions hold $\forall i\in \mathcal V/ \{1\}$ 
			\begin{enumerate}
				\item $d_i(t)$ remains positive and $\phi_i(t)$ and the control input $u_i$ are bounded $\forall t\ge 0$;
				\item the cascade system with state $(\tilde p_2, \tilde v_2, \ldots,\tilde p_n,\tilde v_n)$ has an AS equilibrium point at the origin.
			\end{enumerate}
		\end{prop}
	Proof of this Proposition is provided in Appendix \ref{app:P1}.

		\section{Simulation Results}\label{sec:sim}
\begin{figure}[!htb]
			\centering
			\includegraphics[scale = 0.54]{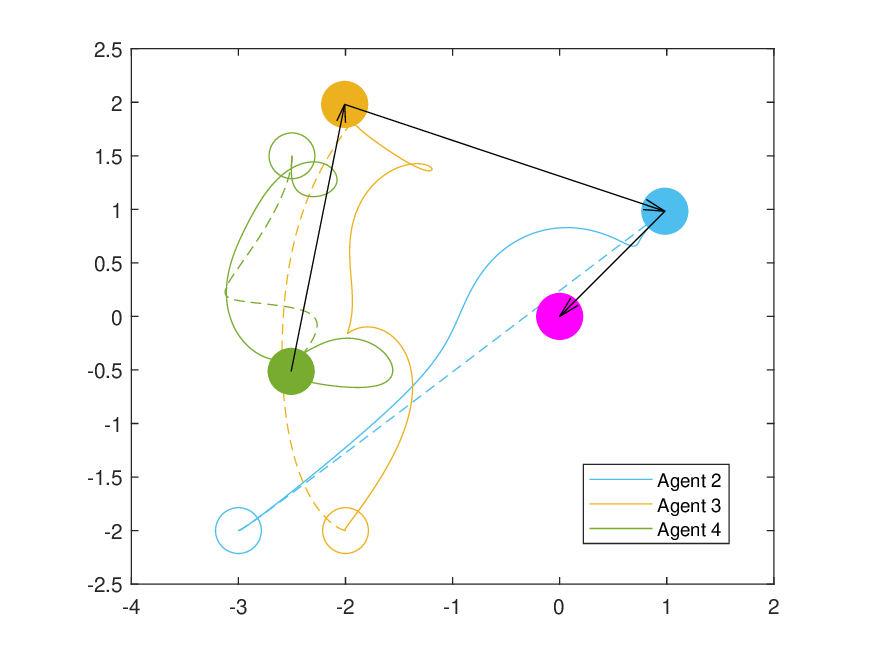}	
			\caption{Trajectories of a leader-follower formation. Solid lines represent trajectories under the controller \eqref{eq:ui} with nominal controller \eqref{eq:u_nd}. Dashed lines represent trajectories under only nominal controllers \eqref{eq:u_nd} without collision avoidance (i.e., $k_{o}=0$). Void and Solid disks are the initial and final positions of the agents, respectively.}
			\label{fig:trajectories_4}
		\end{figure}

\begin{figure}[!htb]
			\centering
			\includegraphics[scale = 0.54]{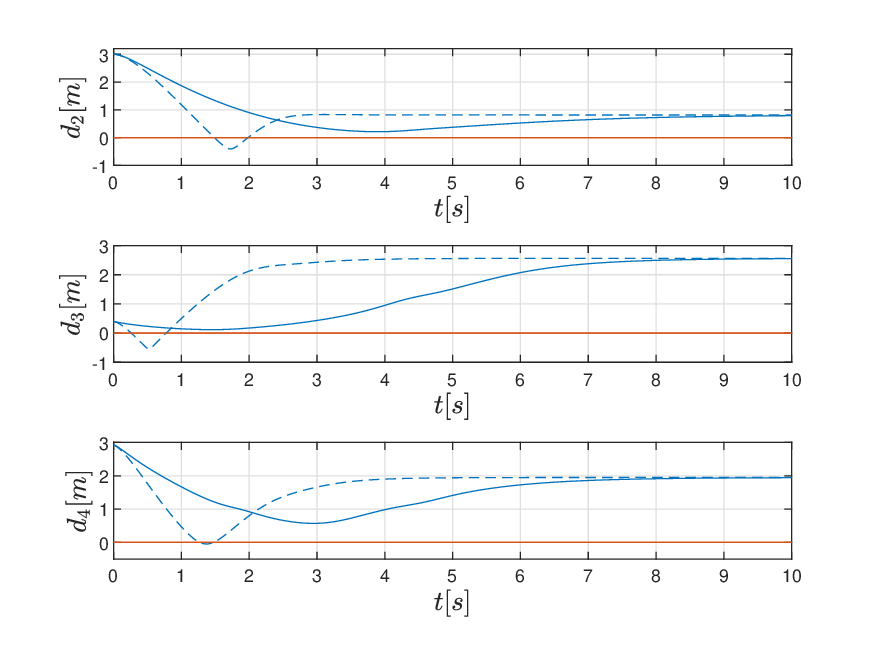}	
			\caption{Evolution of the relative distance $d_i$. Blue solid lines indicate the evolution of $d_i$ under the controller \eqref{eq:ui} with nominal controller \eqref{eq:u_nd}. Blue dashed lines represent the evolution of $d_i$ under only nominal controllers \eqref{eq:u_nd} without using the divergent flow. }
			\label{fig:d_4}
		\end{figure}

\begin{figure}[!htb]
			\centering
			\includegraphics[scale = 0.54]{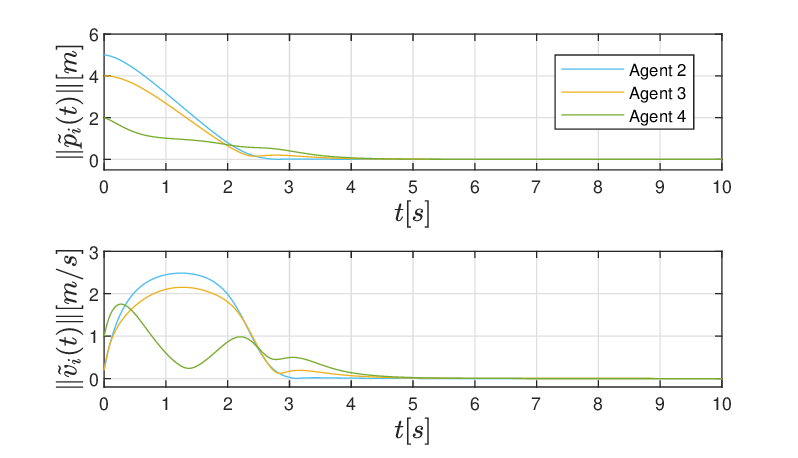}	
			\caption{Time evolution of the state errors.}
			\label{fig:error_4}
		\end{figure}

  		This section provides simulation results to illustrate the effectiveness of the proposed control law for a four-agent system. As indicated in Fig.~ \ref{fig:trajectories_4}, the formation converges to a static desired formation (i.e., $\Omega_i^*=0, \forall i\ge 2$). The gains are chosen as $k_p=10, k_{p_3}=14, k_{p_4}=16, k_v=k_{o_2}=7, , k_{v_3}=k_{o_3}=11, k_{v_4}=k_{o_4}=12$, the initial velocity is  $\boldsymbol v (0)=[(0 \ 0\ 
		0)^\top, (0.2\ 0 \ 0)^\top, (-0.2\ 0 \ 0)^\top, (0 \ -1 \ 0)^\top]$, and the initial positions are indicated in Fig.~ \ref{fig:trajectories_4}. The choices of $h(.)$ in the nominal controller \eqref{eq:u_n} is chosen as $h_{p_i}(.)=1/\sqrt{1+\|\tilde{e}}_i\|^2$ and $h_{v_i}(.)= 1 /{\sqrt{1+\|\tilde{\nu}}_i\|}$.
		The time evolution of the relative distance between neighboring agents is shown in Fig.~\ref{fig:d_4}. Under only the nominal formation controller \eqref{eq:u_n} (that is $k_{o}=0$), Fig.~\ref{fig:trajectories_4} and Fig.~\ref{fig:d_4} show that agents collide with their neighbor. In contrast, the use of the divergent flow in the control law successfully prevents collisions between neighbors (since $d_i$ remains positive as shown in Fig.~ \ref{fig:d_4}) without altering the performance of the nominal controller, as indicated in the Fig.~ \ref{fig:error_4}. To better illustrate the collision avoidance performance, the animation associated  to this simulation can be found in \url{https://youtu.be/3XUA06-Azuo}.
		\section{Conclusion}
  The proposed constructive control design enables a group of second-order dynamic robots to track a desired formation in three-dimensional space while avoiding collisions with neighboring robots. The method utilizes the divergent flow as a barrier feedback added to a nominal formation tracking controller that stabilizes the formation tracking error using relative state variables. Future work includes 1) the generalization of the results for more general graph topologies, including time-varying or switching topologies, and 2) the use of additional information in the nominal controller to reduce the complexity of analyzing the instability cases explicitly.
		\appendix
	\subsection{Proof of Lemma \ref{lem:boundness of OF}} \label{app:L1}
	\begin{proof}
		\textit{ Proof of item 1):}
			Integrating both sides  of equation  \eqref{lem:boundness of OF}, one gets:
			\\
   \begin{equation}\label{eq:int_d}
				k_o (\ln d(t)-\ln d(0))=-(\dot d(t)-\dot d(0))-\int^t_0 \alpha(\tau) d\tau,
			\end{equation}
			a valid relationship as long as $d(t) >0$.
			
			To prove that $d(t)$ will not approach zero in finite time, we use a proof by contradiction.
			Assume that $d(t)$ approaches zero in finite time $T$, that is, $d(T)=0$.  This implies that the left-hand side of equation \eqref{eq:int_d} tends to 'negative' infinity. In contrast, the right-hand side is either 'bounded' or 'positive infinity' because $\alpha(t)$ is bounded, and $\dot d$ is either bounded or negative infinity, which yields a contradiction.
			
			\textit{Proof of item (2):}
			Consider first the forward implication: 
			
			From the above discussion, one concludes that $\lim_{t\to \infty}\int_{0}^t \alpha (\tau)d \tau=+\infty$ as $d$ approaches zero when $t$ tends to infinity.
			
			For the backward implication,  we use a proof by contradiction,  analogous to the proof of item 1). Assume that $d$ is not converging to zero, and $d$ is  positive, either  bounded or tends to infinity as $t \to \infty$. This implies that the left-hand side of \eqref{eq:int_d} is bounded or tends to 'positive' infinity while the right-hand side is 'negative' infinity because  $\lim_{t\to \infty}\int_{0}^t \alpha (\tau)d \tau=+\infty$, $\dot d$ is either bounded or positive infinity. Hence, it yields the contradiction. 
			
			
			\textit{Proof of item (3)}
			
			To show that $\dot d$ converges to zero as $d$ converges to zero when $t$ tends to infinity, we consider, instead of $t$, the new time index $s=\int_0^t \frac 1 {d(\tau)}d\tau$ that tends to infinity iff $t$ tends to infinity.
			
			By rewritten \eqref{ddot_d} as follows:
			\begin{equation}\label{eq:s}
				\frac{d}{ds} \dot d= -k_o\dot d + o(s),
			\end{equation}
			with $o(s)=d\alpha$ a perturbation term that tends to zero when $s$ goes to infinity (or equivalently when $d$ converges to zero). One recognizes the dynamics of a first-order system (with $\dot d$ as the state) perturbed by a vanishing perturbation. From there, one concludes that $\dot d$ is bounded and converges to zero too.
			To prove that $\phi=\frac{\dot d}{d}$ is bounded when $(d,\dot d)\to (0,0)$, we differentiate $\frac{\dot d}{d}$ with respect to $s(t)$. One verifies that:
			\begin{equation}\label{eq:ss}
				\frac{d}{ds} \frac{\dot d}{d}= -(\dot d +k_o)\frac{\dot d}{d}-\alpha,
			\end{equation}
			Using the fact that $\alpha$ is bounded and $\dot d$ is converging to zero, one ensures that there exists a time $T$, such that $(\dot d +k_o)>0, \; \forall t \geq T$. From there, one guarantees that $\frac{\dot d}{d}$  is ultimately bounded by $|\max(\alpha)|/k_o$.

			As for the case when $\alpha(t) \to \alpha^0>\epsilon >0$ and $d \to 0$, one can rewrite $\alpha=\alpha^0 +o(t)$, with $o(t)$ a bounded vanishing perturbation ($o(t)\to 0$). By again analysing \eqref{eq:ss}, one ensures that  $\frac{\dot d}{d} \to -\alpha^0/k_o$. From there, and since $o(t) \to 0$ and $\dot d \to 0$, one concludes  that $\ddot d \to 0$ as $t \to \infty$.
	\end{proof}

\subsection{Proof of Lemma \ref{lem:2agent}} \label{app:L2}
\begin{proof}
 We will first show that the tracking error $(\tilde e_{2},\tilde \nu_{2})$ is bounded. Recalling \eqref{eq:double integrator}, \eqref{eq:ui} and \eqref{eq:u_n} and, using the fact (Assumption \ref{ass:desired}), that $p_1=p_1^*$ and $v_1=v_1^*$, the closed-loop system for the dynamics of the error variable $(\tilde e_{2}, \tilde \nu_{2})$ is expressed as
\begin{equation}\scalebox{0.9}{$ \label{eq:states_21}
		\left\{
		\begin{aligned}
			\dot{\tilde{e}}_{2}&=\Omega_{2}^* \tilde e_{2}+\tilde {\nu}_{2}\\
			\dot{\tilde {\nu}}_{2}&=-\Omega_{2}^*\tilde \nu_{2}-k_{p}h_{p_2}(.)\tilde e_{2}-k_{v}h_{v_2}(.)\tilde \nu_{2}-k_{o}g_{2}\phi_{2}.
		\end{aligned}
		\right.$}
\end{equation}
with 
$\phi_{2}=\frac{\dot d_{2}}{d_{2}}=\frac{g_{2}^\top \nu_{2}}{d_{2}}$. Using the fact that $g_{2} \perp \Omega^*_2 e_{2}=0$, one verifies that  $\phi_{2}=\frac{g_{2}^\top \tilde \nu_{2}}{d_{2}}$.
Consider the following Lyapunov function candidate:
	\begin{equation} \label{eq:L2}
	\mathcal{L}_{2}=\frac{k_p} 2\int_0^{\|\tilde e_{2}\|^2}h_{p_2}(s)ds+\frac 1 2 \|\tilde \nu_{2}\|^2
\end{equation}
One verifies that
\begin{equation}\label{eq:dot L2}
	\begin{aligned}
		\dot {\mathcal{L}}_{2}=-k_{v}\|\tilde \nu_{2}\|^2-k_{o}\dot{d}^2_{2}/d_{2} \leq 0,
	\end{aligned}
\end{equation}
if $d_{2}$ is positive. This indicates that $(\tilde e_{2}, \tilde \nu_{2})$ is bounded  as long as $d_{2}$ remains positive.

{\it Proof of item 1):}
To prove that $d_{2}(t)$ will not approach zero in finite time,  we recall \eqref{eq:states_21} and use the fact that $\dot d_{2}=g_{2}^\top \nu_{2}=g_{2}^\top \tilde \nu_{2}$, to show that:
\begin{equation}\label{eq:ddot_d21}
	\ddot d_{2}=-k_{o}\phi_{2}-k_{v}h_{v_2}(.) \dot d_2 -\alpha'_{2}.
\end{equation}
with $\alpha'_{2}=g_{2}^\top(k_{p}h_{p_2}(.) {\tilde{e}_{2}}-\Omega_{{2}}^{*^2}e_{2})-\frac{\|\pi_{g_{2}}\nu_{2}\|^2}{d_{2}+r}$.  
From \eqref{eq:L2}, it straightforward to verify that all components of $\alpha'_{2}(t)$ are bounded as long as $d_{2}>0$. When $d_{2}=0$ one observes that  the first term in the expression of  $\alpha'_{2}$ is bounded since, in that case, $|e_{2}|=r$ and $\Omega^*$ and $e_{2}^*$ are bounded by assumption. In contrast the second term of $\alpha'_{2}$ is negative and possibly unbounded when $d_{2}=0$. This implies that $\alpha'_{2}$ is either bounded and well-defined $\forall d\ge 0$ or negative and unbounded when $d_{2}=0$. 

Now, by integration both sides of \eqref{eq:ddot_d21}, one gets: 
\begin{equation} \label{eq:integral_d21O} \scalebox{0.9}{$
		k_v\int^t_0 h_{v_2}(.) \dot d_2 d\tau  +k_o \ln \frac{d_{2}(t)}{d_{2}(0)}=-(\dot d_{2}(t)-\dot d_{2}(0))-\int^t_0 \alpha'_{2}(\tau) d\tau $}
\end{equation}
By exploiting condition \eqref{eq:hs} on $h_{v_2}(.)$, one ensures that there exist a positive and bounded scalar $\bar{k}_v(t) \in [k_v \epsilon, k_v \eta]$, such that :
\scalebox{0.9}{$
		k_v\int^t_0 k_{v}h_{v_2}(.) \dot d_2 d\tau =\bar{k}_v(t)(d_{2}(t)-d_{2}(0)) $} and hence, one can rewrite \eqref{eq:integral_d21O} as follows:
\begin{equation} \label{eq:integral_d21} \scalebox{0.9}{$
		\bar{k}_v(t)(d_{2}(t)-d_{2}(0)) +k_o \ln \frac{d_{2}(t)}{d_{2}(0)}=-(\dot d_{2}(t)-\dot d_{2}(0))-\int^t_0 \alpha'_{2}(\tau) d\tau $}
\end{equation}
Analogously to the discussion made after \eqref{eq:int_d}, if there exists a finite time $T>0$ such that $d_{2}(T)=0$, then the left-hand side of equation \eqref{eq:integral_d21} tends to 'negative' infinity while the right-hand side is either 'bounded' or 'positive infinity' because $\dot d_{2}$ and $\alpha'_{2}$ are either bounded or negative infinity, which yields a contradiction. From there, one verifies that the new variable \begin{equation} \label{eq:alpha21}
	\alpha_{2}=k_{v}h_{v_2}(.) \dot d_2 +\alpha'_{2},
\end{equation}
is a bounded continuous function and hence direct application of Lemma \ref{lem:boundness of OF} concludes the proof of Item 1. 


\noindent {\it Proof of Item 2):}  Since $\dot{\mathcal{L}}_{2}(t)\le 0,\forall t>0$, $\mathcal{L}_{2}$ is bounded.
Now, let us analyze the second derivatives of the function $\mathcal{L}_{2}$
\begin{equation} \label{eq:ddotL2}
	\scalebox{0.9}{$
		\begin{aligned}
			\ddot {\mathcal{L}}_{2}&=-2k_{v}\tilde \nu_{2} ^\top \dot \nu_{2} -k_{o}\left(2\ddot d_{2}\phi_{2}-\dot{d}_{2}\phi_{2}^2\right)
		\end{aligned}
		$}
\end{equation}
which is bounded due to the boundedness of $\tilde \nu_{2}$, $\phi_{2}$, $\dot \nu_{2}$, $\ddot d_{2}$. This implies that $\dot {\mathcal{L}}_{2}$ is uniformly continuous, and hence direct application of Barbalat's Lemma implies that $\dot {\mathcal{L}}_{2}$ converges to zero and so does $\tilde \nu_{2}$. From there, one concludes that $\mathcal{L}_{2}(t)$ converges to a constant and $\dot {\tilde e}_{2}\to \Omega_{2}^*\tilde e_{2}$. 



To explicitly derive  $\lim_{t\to \infty} \tilde e_{2}(t)$,  we consider two cases and analyze  \eqref{eq:states_21} and
\eqref{eq:ddot_d21} in sequence.\\
1) First, when $d_{2}(t)>\epsilon>0, \forall t$, a direct application of Barbalat's lemma on \eqref{eq:states_21}, ensures that $\tilde e_{2}$ converges to zero as $\tilde \nu_{2}$ and $\dot d_{2}$ converge to zero guaranteeing that the desired equilibrium point $(\tilde{e}_{2},\tilde{\nu}_{2})=(0,0)$ is asymptotically stable. 
	
\noindent 2) As for the case when $d_{2}\to 0$ as $\tilde \nu_{2}\to 0$, one ensures that the continuous and bounded function $\alpha_{2}$ \eqref{eq:ddot_d21}\footnote{$\alpha_{2}$ exhibits uniform continuity as well, given that all components of $\dot{\alpha}_{2}$  remain bounded. Nevertheless, to maintain a broader scope, we will omit the specific details regarding uniform characteristics.} converges to $k_p\lim_{t\to \infty} g_{2}^\top \tilde e_{2}=\alpha_{2}^0$ which is a constant. By combining the fact that $d_{2} \to 0$ and $\alpha_{2} \to \alpha_{2}^0$, direct application of Lemma \ref{lem:boundness of OF} - items 1) and 3) shows that $\alpha_{2}^0$ is a positive constant and hence one ensures that $\ddot d_{2}$ is bounded and converges to zero as $t \to \infty$.
	
	
	
	Rewrite now the expression $\dot {\tilde \nu}_{2}$ \eqref{eq:states_21} as $\dot{\tilde \nu}_{2}=a_{2}(t)+b_{2}(t)$, with:
	\begin{equation*}\label{eq:dotv21}
		\scalebox{0.9}{$
			\begin{aligned}
				a_{2}(t)&= \pi_{g_{2}}\dot{\tilde v}_{2}=-\pi_{g_{2}}[\Omega_{2}^*\tilde \nu_{2}+k_{p}\tilde e_{2}+k_{v}\tilde \nu_{2}]\\
				b_{2}(t)&=g_{2}g_{2}^\top\dot{\tilde v}_{2}=-g_{2}[g_{2}^\top(\Omega_{2}^*\tilde \nu_{2}+k_{p}\tilde e_{2}+k_{v}\tilde \nu_{2}) +k_{o}\phi_{2}].
			\end{aligned}$}
	\end{equation*}
	From the above discussion, one ensures that $b_{2}(t)$ is bounded and converges to zero. Now, since $a_{2}(t)$ is uniformly continuous because all components of $\dot{a}_{2}$ are bounded, direct application of the slightly different Barbalat's Lemma\footnote{The case $b_{2}(t)\equiv0$, correspond to the classical Barbalat's lemma.} \cite{Claude93} ensures that 
	$\dot{\tilde \nu}_{2}\to -k_{p}\pi_{g_{2}}\tilde{e}_{2} \to 0$. This implies that $\pi_{g_{2}}\tilde{e}_{2}=-\pi_{g_{2}}{e}^*_{2}\to 0$ and $g_{2}\to\pm g_{2}^*$. From there and since $d_{2} \to 0$ one verifies that $e_{2}\to\pm r g_{2}^*$ and $k_{p}  g_{2}^\top\tilde{e}_{2}$ converges to a positive constant and hence, one concludes that $e_{2}\to-rg_{2}^*$ and $\tilde{e}_{2} = e_{2} - e_{2}^* \to -c_{2}^0g_{2}^*$, with $c_{2}^0=(r+c_{2})$. Therefore, the second set point to which $(\tilde{e}_{2}, \tilde \nu_{2})$ may converge is $( -c_{2}^0g_{2}^*,0)$ (as shown in Fig. \ref{fig:2agent} ).
	\begin{figure}[!htb]
		\centering
		\includegraphics[scale = 0.5]{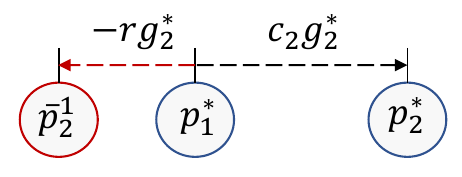}	
		\caption{Possible convergence of agent 2 under the controller \eqref{eq:ui} with nominal control law \eqref{eq:u_n}. The red circle indicates the unstable set point, and the blue one indicates the AS equilibrium point.}
		\label{fig:2agent}
	\end{figure}
	
	To see now that the set point $(\tilde{e}_{2},\tilde \nu_{2})=(\tilde{e}_{2}^0,0)$ (with $\tilde{e}_{2}^0=-c_{2}^0g_{2}^*$) is unstable, recall first that at this point $\|\tilde{e}_{2}\|^2 = (r+c_{2})^2$
	and the Lyapunov function \eqref{eq:L2} is $\mathcal L_{2}^0:=\frac{k_p} 2\int_0^{\|\tilde e_{2}^0\|^2}h_{p_2}(s)ds+\frac{1}{2}\|0\|$.
	
	Denote $R(\epsilon)=\exp(\epsilon \Omega) \in SO(3)$ with $\Omega$ any bounded matrix in the lie algebra  $\so(3)$ such that $\Omega g^*_{2} \neq 0$ (or equivalently $g^*_{2} \notin \ker(\Omega)$). Consider the point $\tilde{e}_{2}^\epsilon=(e_{2}^\epsilon-e_{2}^*) = -r R(\epsilon)g_{2}^* -c_{2} g_{2}^*$ such that $\|\tilde{e}_{2}^\epsilon\|^2 = r^2+{c_{2}}^2 + 2 r c_{2} ({g_{2}^*}^\top R(\epsilon) g_{2}^*)$. Let $\mathcal L_{2}^\epsilon$ denote the value of the Lyapunov function at $(\tilde{e}_{2}^\epsilon,0)$ such that $\mathcal L_{2}^\epsilon=\frac{k_{p}} 2\int_0^{\|\tilde{e}_{2}^\epsilon\|^2}h_{p_2}(s)ds+\frac{1}{2}\|0\|$. Since for any small and positive $\epsilon$, implying that $|{g_{2}^*}^T R(\epsilon) g_{2}^*|<1$ and hence $\|\tilde{e}_{2}^{\epsilon}\|<\|\tilde{e}_{2}^{0}\|$, one concludes that there exist perturbations $R(\epsilon)$ in any neighbour of $(\tilde{e}_{2},\tilde \nu_{2})=(\tilde{e}_{2}^{0},0)$ such that $\mathcal L_{2}^\epsilon <\mathcal L_{2}^0$
	and points infinitely close to $(\tilde{e}_{2},\tilde \nu_{2})=(\tilde{e}_{2}^{0},0)$ converge to $(\tilde{e}_{2},\tilde \nu_{2})=(0,0)$.
\end{proof}
\subsection{Proof of Theorem \ref{thm:control_c}} \label{app:T1}
\begin{proof}
{\it Proof for items 1) $\&$ 2):} 
Recalling \eqref{eq:double integrator}, and \eqref{eq:ui} along with \eqref{f} and \eqref{eq:u_n}, the system dynamics of the error variable $(\tilde e_{i}, \tilde \nu_{i})$, one verifies that: 
\begin{equation}\scalebox{1}{$ \label{eq:states_ii-1}
		\left\{
		\begin{aligned}
			\dot{\tilde{e}}_{i}&=\Omega_{i}^* \tilde e_{i}+\tilde {\nu}_{i}\\
			\dot{\tilde {\nu}}_{i}&=-\Omega_{i}^*\tilde \nu_{i}-k_{p_{i}}\tilde e_{i}-k_{v_{i}}\tilde \nu_{i}-k_{o_{i}}g_{i}\phi_{i}.
		\end{aligned}
		\right.$}
\end{equation}
From there, the proof of items 1) and 2) seamlessly follows from Lemma \ref{lem:2agent}. 
\begin{figure}[!htb]
	\centering
	\includegraphics[scale = 0.5]{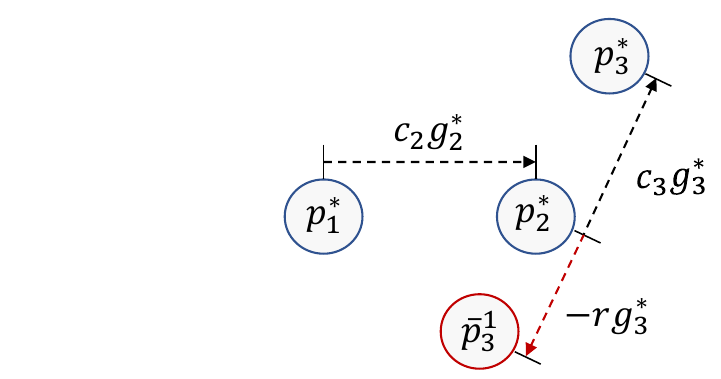}
	\caption{Possible convergence of agent 3 under the controller \eqref{eq:ui} with nominal control law \eqref{eq:u_n} or \eqref{eq:u_nd}, when agent 2 converges to the desired equilibrium point. The red circle indicates the unstable set point, and the blue one indicates the AS equilibrium point.}
	\label{fig:3agent}
\end{figure}
\begin{figure}[!htb]
	\centering
	\includegraphics[scale = 0.5]{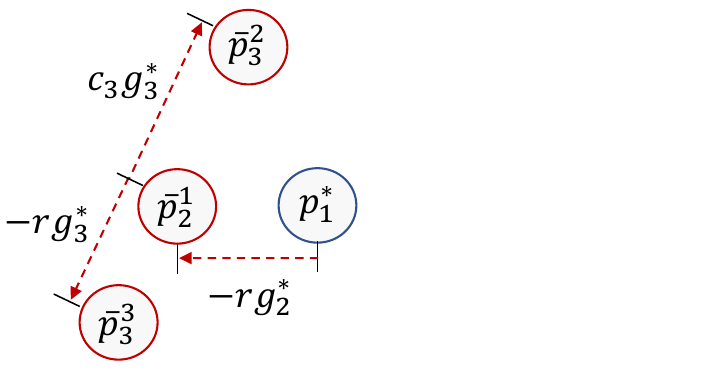}	
	\caption{Possible convergence of agent 3 under the controller \eqref{eq:ui} with nominal control law \eqref{eq:u_n}, when agent 2 converges to the unstable set point. The red circle indicates the unstable set point.}
	\label{fig:3agentB}
\end{figure}

{\it Proof for item 3):} Since $(\tilde e_{i}, \tilde \nu_{i}), 2\le i\le n$ converges towards a binary set point (the equilibrium (0,0) or the unstable set point $(-(r+c_{i}) g_{i}^*, 0)$), one directly concludes that $(\tilde{p}_{i}, \tilde{v}_{i})$ converges to $2^{i-1}$ set points $(\bar {p}_i^m-p_i^*,0), m\in \{1,\ldots, 2^{i-1}\}$, $\forall i\in \mathcal V/ \{1\}$. 
From there, one verifies that the unique asymptotically stable point is $\bar p_{i}^1=p_{i}^*$, $\forall i\in \mathcal V/ \{1\}$ which corresponds to $(\tilde e_{i}, \tilde \nu_{i})=(0,0), \;\forall i\in \mathcal V/ \{1\}$. The remaining $2^{i-1}-1$ set points are by definition unstable since they correspond to at least one unstable point  $(\tilde e_{j}, \tilde \nu_{j})=(-(r+c_{j}) g_{j}^*, 0)$, $j\in\{2,\ldots,i\}$. Figs.~\ref{fig:3agent} and \ref{fig:3agentB} illustrate possible convergence situations of a 3-agent system. 
		\end{proof}

\subsection{Proof of Proposition \ref{thm:control}} \label{app:P1}
	\begin{proof}
	Recalling \eqref{eq:double integrator}, \eqref{eq:ui} and \eqref{eq:u_nd}, one verifies that the closed-loop dynamics of the error variable $(\tilde e_i, \tilde \nu_i), i\ge 2$ can be rewritten as
	\begin{equation}\label{eq:states_i}
		\left\{
		\begin{aligned}
			\dot{\tilde{e}}_i=&\Omega_{i}^*\tilde e_i+\tilde {\nu}_i\\
			\dot{\tilde \nu}_i=&-\Omega_{i}^*\tilde \nu_i-k_{p}h_{p_i}(.)\tilde e_i-k_vh_{v_i}(.)\tilde \nu_i-k_{o}g_i\phi_i\\
			&-(u_{i-1}-u_{i-1}^*)
		\end{aligned}
		\right.
	\end{equation}
	which can be considered as a cascaded system with $(\tilde e_{i-1},\tilde \nu_{i-1})$ perturbing the unforced system \eqref{eq:states_ii-1}.	Analogously to the Proof of Lemma \ref{lem:2agent}, we conclude that the state $(\tilde e_i,\tilde \nu_i)$ of the unforced system \eqref{eq:states_cl} is bounded as long as $d_{i}>0$.
	
	\textit{Proof of Item (1):} Recalling that $\dot d_{3}=g_{3}^\top \nu_3$, from the nominal system \eqref{eq:states_i}, the dynamics of $d_{3}$ can be expressed as 
	\begin{equation}\scalebox{1}{$
			\begin{aligned}\label{eq:ddot_d3j}
				\ddot d_{3}=-k_{o}\phi_3-k_{v}h_{v_3}(.)\dot d_{3}-\alpha^{'}_3, 
			\end{aligned}
			$}
	\end{equation}
	where $\alpha_3^{'}(t)=g_{3}^\top (k_{p_3}h_{p_3}(.)\tilde e_3-\Omega_{i}^{*^2}e_3+u_{2}^*-u_{2})+\frac{\|\pi_{g_{3}}\nu_3\|^2}{d_{3}+r}$. 
	Since $u_{2}$ is bounded (as shown in Lemma 2), the statement of item (1) for $i=3$ can be proved using a similar argument as in Lemma \ref{lem:2agent} - item (1).
	
	Now, if we assume that item (1) is true for $i-1\geq3$, which implies $u_{i-1}$ is bounded, we can conclude analogously to $i=3$, that item (1) is true for $i$. By mathematical induction, one concludes that this item holds for all $i\in \mathcal V/ \{1\}$.
	
	
	{\it Proof of Item (2):} It follows from Lemma \ref{lem:2agent} and Theorem 1 that the unperturbed system \eqref{eq:states_cl} converges either to the AS equilibrium point at $(\tilde e_i, \tilde \nu_i) = (0,0)$ or $(\tilde e_i, \tilde \nu_i) = (-(r+c_i)g_i^*,0)$. By similar arguments, the same will happen for \eqref{eq:states_i} if the perturbation $u_{i-1}$ converges to $u_{i-1}^*$. Thus, one can conclude that the cascaded system with state $(\tilde e_2, \tilde \nu_2, \ldots, \tilde e_i, \tilde \nu_i)$  has an AS equilibrium point at the origin and any remaining set point is unstable. Applying the appropriate change of variables to obtain the state $(\tilde p_2, \tilde v_2, \ldots, \tilde p_n, \tilde v_n)$ yields the claim in Item 2).

\end{proof}

		\addtolength{\textheight}{-12cm}   
		


		%
		%
		%
		%
		%
		%
		%

		\bibliography{bibliography}
		\bibliographystyle{IEEEtran}

	\end{document}